\documentclass[aps,pre,reprint,floatfix]{revtex4-1}
\usepackage{graphicx}
\usepackage{mathpazo}
\usepackage{amsmath}

\newcommand{\sgn}{\operatorname{sgn}}

\begin{document}

\title{Collective Order Decouples Boundary Selection from Macroscopic Chirality in Confined Active Matter}

\author{N\'estor Sep\'ulveda}
\email{nesepulv@gmail.com}
\affiliation{School of Engineering and Sciences, Universidad Adolfo Ib{\'a}{\~n}ez, Diagonal las Torres 2640, Pe\~{n}alol\'en, Santiago, Chile}

\begin{abstract}
Can a boundary determine the direction of circulation in a collectively moving active fluid? We study chiral self-propelled particles with nematic bulk alignment, confined by walls whose orientational coupling varies continuously from nematic to polar. For an isolated particle at a wall, we derive the exact parameter condition under which a stable orientation is tangent to the wall. With predominantly nematic wall coupling, however, the many-body circulation changes sign at a chirality that is nearly independent of the wall symmetry. This sign change follows a pronounced decrease in nematic order and disappears when bulk alignment is removed. As the polar wall coupling becomes dominant, opposite-handed circulation is suppressed and same-handed polar locking emerges. Thus local orientation at a wall and the direction of macroscopic circulation are selected by distinct physical processes: the wall determines the possible local orientations, whereas collective order determines whether those orientations control the global flow.
\end{abstract}

\maketitle

\section{Introduction}

Active matter systems, collections of self-propelled units that convert stored energy into directed motion, generate collective states with no equilibrium counterpart~\cite{marchetti2013hydrodynamics, Gompper2025, Vicsek1995}. A recurring, still incompletely understood question is how \emph{chirality} propagates across scales: when individual units carry an intrinsic rotational bias, does the resulting collective flow rotate in the same sense, the opposite sense, or not at all? Confinement can organize otherwise disordered active particles into macroscopic vortices~\cite{Bricard2015}, and collective polar vortices can retain and reverse their selected handedness~\cite{Zhang2022}. Edge flows with a well-defined circulation sense have been reported in bacterial colonies~\cite{PhysRevX.14.041006}, while confined chiral particles develop wall-parallel currents and redistribute between boundary and bulk~\cite{CapriniMarconi2019,Cruz2024}. Ordered chiral-fluid theories further show that intrinsic rotation can stabilize orientational order~\cite{MaitraLenz2019}, and experiments on anisotropic rollers display reconfigurable polar and nematic chiral states~\cite{Zhang2020}. An exact equation of state relating edge flux to bulk odd stress has recently been established for chiral colloidal fluids~\cite{Metzger2026}, while a microscopic hydrodynamic derivation connects chirality-induced torque, odd viscosity, and edge currents~\cite{MarconiCaprini2026}. In each case the sign relationship follows from the specific microscopic mechanism and boundary symmetry, not a universal rule.

A second, largely independent axis of variation is the symmetry of the alignment interaction: \emph{polar} systems align head-to-tail~\cite{Vicsek1995}, while \emph{nematic} systems align along a common axis without a preferred direction, as in confluent epithelial monolayers, where it is associated with confinement phenomena such as spontaneous shear flow~\cite{Duclos-Silberzan2018} and bidirectional laning~\cite{Lacroix-Silberzan2024}, mediated by collective organization near the boundary. Real boundaries can couple to orientation through mixtures of polar and nematic contributions: a wall, substrate cue, or channel edge need not be purely one symmetry or the other. The natural expectation is that this mixture, together with the sign of the intrinsic chirality, fixes the sign of the resulting macroscopic circulation -- a single, wall-encoded rule.

Here we separate two questions that are usually conflated: which orientations are stable for an isolated particle at a wall, and which direction of circulation is selected by the interacting suspension. We define the orientational coupling relative to the outward normal of each wall and vary its symmetry continuously from nematic to polar. This construction gives an exact, parameter-free condition for a stable single-particle orientation to become tangent to the wall. We then compare this local prediction with the chirality at which the many-body current changes sign. With predominantly nematic wall coupling, the many-body sign change is nearly independent of wall symmetry and disappears when bulk alignment is removed. As polar wall coupling becomes dominant, the opposite-handed many-body state is suppressed and the response crosses over to same-handed polar locking.

\section{Model}

We model $N$ self-propelled particles confined between two parallel walls separated by a channel of width $L_\perp$, periodic along the channel axis:
\begin{align}
\dot{\mathbf r}_i &= v_0 \hat{\mathbf u}_i + c_r(\mathbf f_i + \mathbf f_{wi}), \label{eq:pos}\\
\dot\theta_i &= \omega_0 + c_\theta\sum_{j}\sin[2(\theta_j-\theta_i)] + \sigma_\theta\xi_i + \tau_{wi}, \label{eq:ang}
\end{align}
where $\hat{\mathbf u}_i=(\cos\theta_i,\sin\theta_i)$, $\mathbf f_i$ is a short-range Gaussian repulsion, $\omega_0$ is the signed intrinsic angular velocity (positive for counterclockwise rotation), and $\langle\xi_i(t)\xi_j(t')\rangle=\delta_{ij}\delta(t-t')$. The wall torque, for a particle within distance $a$ of wall $W$, is
\begin{equation}
\tau_{wi} = \mathbf 1(d_{iW}<a)\left[-K_p\sin\psi_i + K_n\sin2\psi_i\right],
\label{eq:tau_w}
\end{equation}
with $\psi_i\equiv\theta_i-\theta_{n,W}$ the orientation measured from wall $W$'s own local outward normal $\theta_{n,W}$ ($0$ at the right wall, $\pi$ at the left), so both walls remain related by the problem's reflection symmetry. A bounded mixing parameter sets the wall torque's polar/nematic split at fixed total intensity,
\begin{equation}
K_n = c_\theta(1-\lambda), \qquad K_p = c_\theta\lambda, \qquad \lambda\in[0,1],
\label{eq:lambda}
\end{equation}
interpolating between purely nematic ($\lambda=0$) and purely polar ($\lambda=1$) anchoring. We define the circulation order parameter
\begin{equation}
J = \frac{1}{Nv_0}\sum_i s(x_i)\,v_{y,i}, \qquad s(x)=\sgn(x-x_c),
\label{eq:J}
\end{equation}
with $x_c$ the channel centerline. We call the circulation same-handed when $J$ and $\omega_0$ have the same sign, and opposite-handed when their signs differ. The repulsion in Eq.~\eqref{eq:pos} is the truncated, penetrable Gaussian
\begin{equation}
\mathbf f_i=\sum_{j:r_{ij}<a}e^{-r_{ij}^2/(2\sigma^2)}\hat{\mathbf r}_{ij},
\qquad \sigma=a/2,
\label{eq:repulsion}
\end{equation}
which suppresses clustering without introducing a competing hard-core jammed state; the wall force $\mathbf f_{wi}$ is instead a Hookean linear spring of range $a$, normalized so it is dimensionless like $\mathbf f_i$ -- full form in the SM. Unless noted otherwise, simulations use $N=450$, $L_\perp/a=12$, $L_\parallel/a=30$, and hence $\rho_0a^2=1.25$, with $c_\theta=35.32\,\mathrm{h}^{-1}$, $\sigma_\theta^2/c_\theta=0.25$, a $250\,\mathrm{h}$ equilibration period, and a $48.3\,\mathrm h$ production window. We integrate positions with a Heun predictor--corrector and angles with Lie--Trotter substepping at $dt=1.04\times10^{-3}\,\mathrm h$; configurations are sampled every $80$ steps. Brackets denote production-time and seed averages, and uncertainties are standard errors across seeds after time averaging each realization. With the simulated reference values $a=25\,\mu\mathrm m$ and $v_0=10\,\mu\mathrm m\,\mathrm h^{-1}$, of epithelial-cell order, the convective time is $a/v_0=2.5\,\mathrm h$. In these units, the bulk dynamics is specified by the dimensionless groups $c_r/v_0$, $c_\theta a/v_0$, $\sigma_\theta\sqrt{a/v_0}$, and $\omega_0a/v_0$ (so $c_r$, like $v_0$, carries velocity units, and the Gaussian repulsion kernel $\mathbf f_i$ in Eq.~\eqref{eq:repulsion} is dimensionless), in addition to $\lambda$ and the geometric ratios. Further parameter and protocol details are given in the Supplemental Material (SM).

\begin{figure}[tbp]
  \centering
  \includegraphics[width=\linewidth]{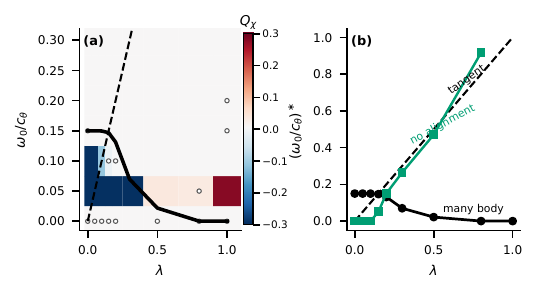}
  \caption{Local wall prediction compared with the many-body current. (a) State map of $Q_\chi=\sgn(\omega_0)\langle J\rangle$ in the $(\lambda,\omega_0/c_\theta)$ plane ($N_{\rm seeds}=18$; color range clipped at $\pm0.3$). Negative and positive values indicate opposite- and same-handed circulation, respectively. Open circles mark points with $|\langle J\rangle|/\mathrm{SEM}\le2$. The dashed curve gives the condition for an isolated particle to be tangent to the wall, Eq.~\eqref{eq:boundary_line}; the solid curve gives the first zero of the mean many-body current after a negative-current interval. (b) Here $(\omega_0/c_\theta)^*$ denotes the characteristic chirality ratio associated with each curve: the first zero of the mean many-body current, the zero of the population-averaged wall orientation without bulk alignment, or the isolated-particle tangent condition. When bulk alignment is removed ($N_{\rm seeds}=15$), the nearly $\lambda$-independent many-body sign change disappears. For $\lambda\gtrsim0.2$, the remaining change in the population-averaged wall orientation occurs near the isolated-particle prediction.}
  \label{fig:1}
\end{figure}

\section{Single-particle wall dynamics}

Near a wall, and neglecting interparticle interactions and noise, Eq.~\eqref{eq:tau_w} reduces $\dot\theta_i=\dot\psi_i$ to
\begin{equation}
\dot\psi = \omega_0 - K_p\sin\psi + K_n\sin2\psi.
\label{eq:psi_dot}
\end{equation}
Equivalently, $\dot\psi=-\partial_\psi U$ in the tilted two-harmonic potential
\begin{equation}
U(\psi)=-\omega_0\psi-K_p\cos\psi+\frac{K_n}{2}\cos2\psi .
\label{eq:potential}
\end{equation}
The nematic torque alone has two tangent equilibria, $\psi=+\pi/2$ ($s=+1$) and $\psi=-\pi/2$ ($s=-1$). Evaluating Eq.~\eqref{eq:psi_dot} at $\psi=s\pi/2$ gives, exactly,
\begin{equation}
\frac{\omega_0}{c_\theta} = s\lambda, \qquad s=\pm1,
\label{eq:boundary_line}
\end{equation}
the condition under which $\psi=s\pi/2$ is a fixed point. Thus the stable orientation is exactly tangent to the wall at this parameter value. For $K_n>0$, the fixed point is linearly stable because $\partial_\psi\dot\psi|_{s\pi/2}=-2K_n<0$. Equation~\eqref{eq:boundary_line} is a local statement about one fixed-point solution: it specifies where that solution changes from one side of the tangent direction to the other. It does not determine which stable solution is populated by the interacting system and therefore does not determine the sign of $J$. This tangent condition is also different from the chirality at which all wall-locked fixed points disappear,
\begin{equation}
\frac{|\omega_0|_{\rm dep}}{c_\theta}
=\max_\psi\left|\lambda\sin\psi-(1-\lambda)\sin2\psi\right|,
\label{eq:depinning}
\end{equation}
where the final stable fixed point merges with an unstable fixed point and both cease to exist. A third parameter value marks the loss of coexistence: one of the two stable solutions disappears while the other remains. For example, at $\lambda=0.3$ two stable solutions coexist only up to $\omega_0/c_\theta\simeq0.49$, whereas the remaining solution persists up to $|\omega_0|_{\rm dep}/c_\theta\simeq0.92$. At $\lambda=0.8$, only one stable solution exists even at zero chirality. At $\lambda=1$, Eq.~\eqref{eq:psi_dot} reduces to $\dot\psi=\omega_0-c_\theta\sin\psi$, and its stable fixed point disappears at $|\omega_0|/c_\theta=1$. The complete list and stability of the fixed points are given in the SM.

\section{Many-body circulation}

\subsection{State diagram}

We map $Q_\chi=\sgn(\omega_0)\langle J\rangle$ over $\lambda\in\{0,\ldots,1\}$ and $\omega_0/c_\theta\in\{0,\ldots,1\}$, with $N_{\rm seeds}=18$ per point (Fig.~\ref{fig:1}). At $\omega_0=0$, $Q_\chi$ is defined to be zero and therefore provides no information about spontaneous circulation. For $\lambda\lesssim0.10$, the mean current changes sign near $\omega_0/c_\theta\approx0.15$, almost independently of $\lambda$ and far from the isolated-particle prediction in Eq.~\eqref{eq:boundary_line}. As $\lambda$ increases, the range with a strong negative current becomes smaller. Combining the central grid with new fine-grid simulations gives first-zero estimates of $0.150,0.150,0.131,0.0699,0.02185$, and $0.000134$ for $\lambda=0,0.10,0.20,0.30,0.50,0.80$, respectively. The last three estimates use 30, 40, and 50 seeds per point. At $\lambda=0.3$, $\langle J\rangle$ remains close to zero after the first zero and changes sign again on the fine grid; $0.0699$ therefore marks the end of the strong negative-current state, not a sharp transition to a finite positive current. At $\lambda=0.8$, the estimated interval is effectively absent. The SM reports bootstrap intervals and classification probabilities. At $\lambda=1$, the mean current is not negative at any positive chirality sampled; the system instead shows strong same-handed circulation and polar orientation at the wall. We therefore distinguish collective control for $\lambda\lesssim0.10$, competition between collective and wall effects at intermediate $\lambda$, and predominantly polar-wall control as $\lambda$ approaches 1. Across all $126$ parameter combinations ($2268$ runs), the largest difference between the mean orientations at the two walls was $0.061$ rad, consistent with finite sampling.

\subsection{Collective origin}

\begin{figure}[tbp]
  \centering
  \includegraphics[width=\linewidth]{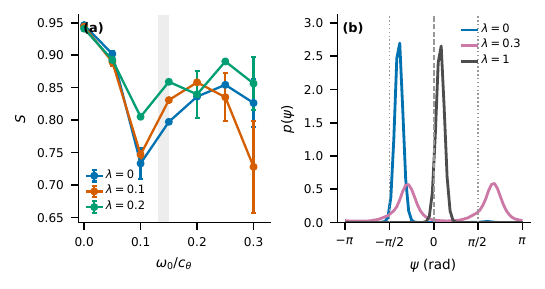}
  \caption{Orientational changes associated with reversal of the mean current. Colors denote the $\lambda$ values listed in each panel; blue denotes $\lambda=0$ in both. (a) Nematic order $S$ for $\lambda=0,0.1,0.2$; its pronounced decrease precedes the sign change of $\langle J\rangle$. (b) Wall-angle distributions at $(\lambda,\omega_0/c_\theta)=(0,0.10),(0.3,0.10),(1,0.05)$. The wall normal is $\psi=0$ and the tangent directions are $\psi=\pm\pi/2$ (dotted).}
  \label{fig:2}
\end{figure}

To test whether the sign change at small $\lambda$ requires collective alignment, we repeat the same parameter grid with particle--particle alignment removed while retaining the wall torque. The sign change near $\omega_0/c_\theta=0.15$ then disappears. For $\lambda\gtrsim0.2$, the chirality at which the mean wall orientation changes sign differs by $5$--$28\%$ from the isolated-particle value in Eq.~\eqref{eq:boundary_line}. For $\lambda\lesssim0.15$, that chirality approaches zero, whereas the many-body current with alignment changed sign near $0.15$ (Fig.~\ref{fig:1}b). This comparison shows that the nearly $\lambda$-independent sign change requires bulk alignment. It is preceded by a pronounced decrease in the nematic order $S=\langle|N^{-1}\sum_i e^{2i\theta_i}|\rangle_t$, defined using the instantaneous modulus before time averaging so that rotation of the director does not artificially reduce $S$ (SM). For $\lambda=0,0.1,0.2$, $S$ decreases from $\approx0.94$ at $\omega_0/c_\theta=0$ to $0.73$--$0.81$ at $\omega_0/c_\theta=0.10$, then partially recovers while $\langle J\rangle$ changes sign at $\omega_0/c_\theta\approx0.13$--$0.15$ (Fig.~\ref{fig:2}a). The decrease in order is therefore associated with the change in circulation but neither determines its exact location nor demonstrates causality. The pooled distribution of orientations near the walls, $p(\psi)$, changes from a narrow peak near a wall-tangent direction at $\lambda=0$ to a narrow peak near the wall normal at $\lambda=1$. At $\lambda=0.3$, the distribution is broad and has only a weak maximum, consistent with competition between these two forms of wall orientation (Fig.~\ref{fig:2}b).

These observations show that bulk alignment is required for the sign change near $\omega_0/c_\theta\simeq0.15$, but they do not provide an analytic prediction for this value. They also distinguish two testable responses. Changing a bulk parameter should alter both the many-body sign change and the associated decrease in nematic order. Changing the polar component of the wall torque instead shifts the isolated-particle tangent condition according to Eq.~\eqref{eq:boundary_line}. The similar decrease in order for the three small-$\lambda$ curves and the result obtained without bulk alignment support this distinction. Deriving the many-body sign-change value from a coarse-grained theory remains an open problem.

To quantify the asymmetry of the near-wall orientations, define $P_w=[\Pr(\psi>0)-\Pr(\psi\le0)]/[\Pr(\psi>0)+\Pr(\psi\le0)]$. These probabilities are calculated from all recorded particle orientations within distance $a$ of either wall. Thus $P_w$ measures which angular half-plane is more populated; it measures the populations of distinct stable solutions only when the single-particle analysis in the SM confirms that two such solutions coexist. At the three points in Fig.~\ref{fig:2}b, $(P_w,\langle J\rangle)$ is $(-0.987,-0.588)$ for $\lambda=0$, $(+0.019,\text{near }0)$ for $\lambda=0.3$, and $(+0.884,+0.274)$ for $\lambda=1$. Stochastic integrations of Eq.~\eqref{eq:psi_dot} for one isolated particle instead give $P_w=-0.001,+0.754,+0.116$, respectively. This comparison includes the angular dynamics at the wall but omits how long a particle remains there, the wall-normal force, and repeated motion into and out of the wall region. It is therefore not a quantitatively equivalent noninteracting reference for the many-particle system. Subject to this limitation, interactions create an angular bias that is absent for an isolated particle at $\lambda=0$, reduce a strong isolated-particle bias at $\lambda=0.3$, and increase a weaker one at $\lambda=1$. Collective organization therefore does not simply multiply the isolated-particle response by a constant factor.

The deterministic control separates basin selection from noise-driven excursions. With random initial angles and $\sigma_\theta=0$, the two coexisting attractors at $\lambda=0$ and $0.3$ receive nearly equal populations ($P_w=-0.002$ and $-0.023$), whereas every $\lambda=1$ trajectory converges to the sole polar fixed point ($P_w=1$). Spatial profiles provide a complementary many-body signature at $\omega_0/c_\theta=0.05$. Near-wall nematic order is high at both endpoints, $S_{\rm wall}=0.964$ for $\lambda=0$ and $0.962$ for $\lambda=1$. At the separate intermediate point $\lambda=0.5$, it falls to $0.785$, below the corresponding bulk value $S_{\rm bulk}=0.925$. This spatial contrast indicates weakened wall locking, rather than cancellation between two equally ordered tangent populations.

\begin{figure}[tbp]
  \centering
  \includegraphics[width=\linewidth]{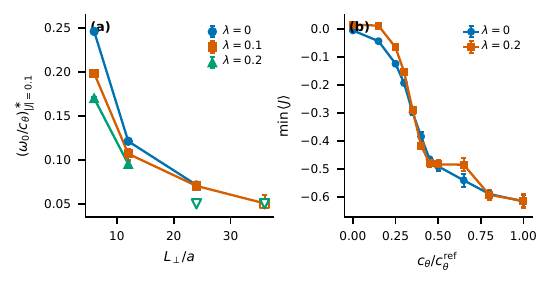}
  \caption{Dependence on confinement and interparticle alignment. (a) Chirality at which $|\langle J\rangle|$ first falls below $0.1$ for $\lambda=0,0.1,0.2$ (seed-bootstrap $68\%$ intervals). The open square has $51\%$ bootstrap resolution; open downward triangles are left-censored at $<0.05$. (b) Most negative sampled $\langle J\rangle$ versus $c_\theta/c_\theta^{\rm ref}$ for $\lambda=0$ and $0.2$ (SEM at the selected grid point). In this control, $c_\theta$ denotes the independently varied interparticle coupling, while the wall torque is fixed; $c_\theta^{\rm ref}=35.32\,\mathrm{h}^{-1}$.}
  \label{fig:3}
\end{figure}

\subsection{Generality, predictions, and scope}

At four times the reference angular-noise intensity, the opposite-handed signal at $\lambda=0$ is no longer resolved, whereas same-handed polar locking at $\lambda=1$ remains resolved at $\omega_0/c_\theta=0.05$. At $\lambda=0$, the collapse threshold decreases from approximately $0.25$ at $L_\perp/a=6$ to $0.07$ at $L_\perp/a=24$ (Fig.~\ref{fig:3}a). Its product with width varies from $1.45$ to $1.73$, suggesting, but not proving, an inverse dependence. New simulations show the same monotonic confinement dependence at $\lambda=0.1$; at $\lambda=0.2$ the response for $L_\perp/a\ge24$ is already below the $|\langle J\rangle|=0.1$ criterion at the smallest sampled chirality. By contrast, changing the periodic length from $L_\parallel/a=10$ to $60$ at fixed width and density leaves the $\lambda=0$ current sign change between $0.144$ and $0.150$. The confinement result is therefore not explained by the total particle number or periodic length alone. An expanded 20-seed scan of bulk alignment shows that the most negative sampled current decreases continuously but nonlinearly from $-0.617$ at the reference coupling to $-0.004$ without alignment (Fig.~\ref{fig:3}b). At $\lambda=0.2$, a negative-current state first becomes resolved between alignment ratios $0.15$ and $0.25$. These controls confirm that sufficiently strong bulk alignment is required, while replacing the earlier inference of an abrupt threshold between ratios $0.25$ and $0.5$ by a gradual nonlinear crossover. Numerical details and the completed-control summary are given in the SM.

\section{Conclusion}

The parameter value at which an isolated particle becomes tangent to a wall does not necessarily determine the direction of the many-body circulation. With predominantly nematic wall coupling, the mean current changes sign at a value that has no isolated-particle counterpart. This behavior persists over the tested channel widths and periodic lengths, for both tested repulsion kernels, and under moderate reductions of bulk alignment; it disappears when bulk alignment is sufficiently weak. Its numerical location depends on density and, near the crossover, on preparation. As polar wall coupling becomes dominant, the opposite-handed state contracts and same-handed polar locking takes over; at $\lambda=1$, no negative mean current is found at any sampled positive chirality. The robust conclusion is that local orientation at a boundary and macroscopic transport can be selected independently.


\bibliography{active}
\end{document}